# The Use of ICT to preserve Australian Indigenous Culture and Language – a Preliminarily Proposal Using the Activity Theory Framework

**Sarah Van Der Meer**
Department of Computing
Faculty of Science and Engineering
Macquarie University
Australia
Email: sarah.vandermeer@mq.edu.au

**Stephen Smith**
Department of Computing
Faculty of Science and Engineering
Macquarie University
Australia
Email: stephen.smith@mq.edu.au

**Vincent Pang**
School of Information Systems, Technology and Management
UNSW Business School
University of New South Wales
Australia
Email: vincent.pang@unsw.edu.au

## Abstract

Propinquity between Australian Indigenous communities' social structures and ICT purposed for cultural preservation is a modern area of research; hindered by the 'digital divide' thus limiting plentiful literature in this field in theoretical or practical applications. Consequently, community consultations become mandatory for deriving empirical and effective processes and outcomes in successful culture and language preservation and teaching of Indigenous culture in Aboriginal Australian communities. Analysis of a literature review has identified ICT as the best provision method to immortalize and teach cultural knowledge and language for Indigenous Australians determined by the accessibility of ICT's, the capacity of Aboriginal Australians to learn to use ICT and in some instances, the increased cost effectivity for multi-community communications and meetings from geographically dispersed land councils to use ICT. This research examines the effectiveness and outputs of culturally conscious, end-user driven ICT development and implementation into contemporary Indigenous Australian social structures and communities.

# 1  Introduction

More than 100 out of the 250 languages spoken by Indigenous people have become extinct since 1788, the arrival of Captain Cook to Australia (Harrington 2014). Consequently, the preservation of the remaining Aboriginal Australian cultures and languages with emphasis on the teaching of endangered Aboriginal Australian cultures and languages is a critically important issue for the Aboriginal Australian communities of Australia. Aboriginal Australian identities are unyieldingly interlinked with their connection and proficiency in cultural knowledge, community engagement and use of language and the land. Unfortunately, without connection to land, knowledge and language, Aboriginal Australians substitute fulfilment from these connections with addictive illicit substances which contributes to the decline of social capital and depreciation of contemporary Indigenous social structures (Amos 2015; Burke 2015; Everingham 2015; Jones 2015; Schubert 2015). This results in hindering capacity to self-actualize as an independent or connected society which is empirically harmful for the wellbeing and continuity of a culture (Anonymous 2015d).

Preservation of Aboriginal Australian culture and language has occurred over multiple platforms inclusive but not limited to, tourist tours and attractions hosted by cultural communities, paintings, art galleries, performances of music, songs, dance and ceremony, national parks and protection of traditional sacred sites (Anonymous 2015; Lawrie 2012), recording of oral history and cultural heritage and attempts to circulate this to younger Aboriginal members of the community (Lawrie 2012). The Ara Iritija Project is one of the most renowned preservation projects and aims to return culturally significant materials to native persons, inclusive of; photographs, films, sound recordings and documents which it then stores on a purpose built computer (Gibson 2009).

The Internet enables the preservation of historical heritages and has acted as a promoter for traditional cultures (Cui and Yokoi 2012). The existences of e-museums and namely the "Online Museum for History … as a collection of digitalized information resources, such as images, textual documents, 3D models, flash documents, videos, and audio files of relics collected in physical museums" (Cui and Yokoi, 2012, p. 2). Moreover, in some instances, number of visitors visit to e-museums has already out-numbered physical (on-site) visitors (Kravchyna and Hastings 2002). The ever-growing popularity and use of e-museums as sources of information suggests the movement of other cultural preservation projects existing more commonly through the use of IS.

There are existing Aboriginal Australian online museums (Anonymous 2015c), however these have been unable to reach their full potential or have failed due to segregated structures of their community information. This mean the data structure aligns with communities but does not span the cultural divide. This stratified data design obliges with community segregation of knowledge sharing rules of the culture, however, without engaging with the contemporary social structures (Anonymous 2015b). These geographic clusters currently cannot be linked into one system due to technical and network constraints.  This is challenging as the literacy skill and IT competency of individuals in these communities is basic. The current systems rely on these communities to fund and maintain technology, currently there is no knowledge platform, this has repressed the success of pre-existing cultural e-museums – and why previous attempts of producing an e-museum have failed. By integrating the concept of segregated information and authorized access to these databases we propose to capture the cultural knowledge with contemporary social structures.

The design for these information systems is invaluable because:

1) Aboriginal Australians will be in control of how the system portrays culture from a strategic level (NSW ALC) and local knowledge level (LALC).

2) It is important for these information systems and databases to belong to Aboriginal Australians as it is about their race and culture – and the best way for these systems to be a part of and belong to Aboriginal Australians is by integration into contemporary social structures.

3) These systems should be integrated into the contemporary social structures because these systems are an innovation of Aboriginal Australians' teaching and sharing of knowledge which is traditionally performed by rights determined by social structures within each 'mob' nation, and as such, continuity of this method of teaching and sharing knowledge should be prevalent to ensure the respectful treatment of knowledge and following of social structural rules where the sharing of knowledge is concerned. Granted this information is collated and shared by communities to teach younger generations of Aboriginal Australians to preserve the Aboriginal

    Australian culture in the most holistic approach possible, there is value to be extrapolated from the knowing that this system belongs to Aboriginal Australians.

4) The purpose of the NSW ALC, RALC's and LALC's amongst other purposes is to ensure the "maintenance and enhancement of Aboriginal culture, identity and heritage (including the management of traditional sites and cultural materials within NSW)" (Anonymous 2015b) and to protect the interests and further the aspirations of its people.

Henceforth, this paper proposes the design and development of an Aboriginal Australian culture/ history preservation/ teaching based information system to act as an online museum with authorization rights. Access to these online museums will be exclusively accessible by individuals/ communities to their respective museum communities/ individuals from one Local Aboriginal Land Council (LALC) cannot access the information of another communities LALC in order to maintain cultural sovereignty for each LALC and maintain the traditional passing of knowledge within individuals of one community and the traditional form of sharing knowledge between LALC's – meeting in person – to ensure the integrity of the knowledge stored is not lost and maintains the traditional treating of knowledge.

Sound developments in the use of Information Communication Technologies (ICT's) to preserve Aboriginal Australian culture and language have been made (Gibson 2009; Bandias 2010; Mohammad and Yi-Chan 2010; Featherstone 2011; McMahon 2011; Madden, et al 2012; Featherstone 2013). The extent of ICT's capacity to preserve Aboriginal Australian's culture is yet to be actualized. Prior hindrances incurred by the 'digital divide' are less imposing than 10 years ago for Aboriginal Australians. Small, individual communal projects to create information systems to record, preserve and access cultural knowledge and language have been created, this papers proposal is to build upon these ideas and approaches – community consultation as one of multiple discusses approaches explored – and roll out a state wide information system that will provide individualistic knowledge, languages and culture corresponding to the respective community which will cater to the 119 individual LALC's in NSW that may 'opt' in to having a community online museum built.

The Literature Review engenders credence that the design, development and operation of ICT's which are governed by relevant Aboriginal Australian epistemologies and ontologies will best preserve and make accessible; cultural knowledge and language in increasingly interactive; in depth ways and additionally enabling future generations of Aboriginal Australians to learn and teach traditional culture and language through their LALC's.

## 2 Literature Review

The extant literature encompasses learning of culture and learning with ICT. Additionally, contrasts and comparisons of traditional and contemporary forms of cultural preservation are reviewed. The impacts of the rollout of the NBN (National Broadband Network) and 3G (Next Gen phone reception) on the digital divide in rural and metropolitan Aboriginal Australian communities with noted emphasis on Aboriginal Australian's youth embracement of ICT's are analysed in regards to the productive and cultural mannerisms in which these ICT's are adopted.

The literature review is structured to address four main themes related to this research:

1. Western researchers prevalent perceptions of the harms of ICT and threats which are posed to Indigenous cultures and the integrity of those cultures as they uptake the use of ICT on a daily basis.

2. Speaking to the previous point, elaboration as to why these perceptions are inaccurate from a system design and implementation perspective.

3. Literature refuting the notion that ICT harms Aboriginal Australian culture's integrity, affirmed by personal interviews and the ready embracement of ICT by Aboriginal Australian communities - notably by the youth of these communities; henceforth closing the digital divide and enabling the greater preservation of culture, namely Ara Irititja Archives and community uses of ICT to generate media and art and music

4. The values in the greater accessibility of the wider global community in delivering opportune prospects of travel, employment and education, henceforth increasing the social capital of Aboriginal Australian individuals and communities provided by ICT uptake.

## 2.1 Western Perceptions of ICT Harms of Adoption by Indigenous Culture

Throughout the review literature there is a prevalent idea that "ICTs have the power to change our culture and transform us and the way we think, then this has serious implications for the adoption of the technology by Indigenous peoples, struggling to maintain the integrity of their culture in a world dominated by Western ideologies and lifestyles" (Dyson 2004)

This notion is echoed by various other academics whom outcry against the uptake of ICT's and exposure of Western culture to Aboriginal Australian communities in aim of mitigating the depletion of the existing culture and to prevent Westernization or hindrance of Aboriginal Australian communities' cultures capacity to thrive as a consequence of ICT inherently embodying Western cultural values (Dyson 2004; Dugdale et al. 2005; Dyson and Underwood 2006; Taylor 2012a; 2012b; Featherstone 2013).

Whilst holding logical merit this perspective establishes a situation of quandary where resolution and prevention is concerned;

1. How should Aboriginal Australian communities and individuals interact with the world while they seek remain an isolated from?
2. Who should manage to preservation of Aboriginal Australian culture? When these Aboriginal Australian communities are forbidden to interact with the media and way of life of the globalised community?

The stance of these academics can also be, and will be debated in so far as, information systems will embody the culture, values and developed for this purpose when designed with community consultation and existing social structures (NSW ALC, RALC's, LALC's) for the use of appropriate resources. This equates, as the instance that an information systems is created with Aboriginal Australian language and values of culture and is purposed for the preservation of Aboriginal Australian culture the system will achieve its objectives and help culture or cultural integrity is foreseeable.

## 2.2 System Design and Implementation Perspective

The Aboriginal Australian culture has both characteristics of individualism and collectivism. It has collectivism characteristic because in each LALC, people work and share knowledge together within that community. On the other hand, the individualism characteristic reflects each LALC not sharing their knowledge with other LALC. No study in IS has yet to address this subject.

Information systems are constructions and embodiments of the cultural values with which they are developed. However in this case they need to be implemented into an Aboriginal Australian information system of knowledge and practice are embedded within language and institutionalized by language. "What is known, how knowledge is gained, how knowledge is defined and expressed is, to a large extent, determined by language and its use in context" (Fogarty and Kral 2011).

Throughout the literature review is a prevalence that ICT's are culturally adaptive and can be appropriated and embody the values of any culture and continue to function. The culturally-adaptive nature of ICT's have and will continue to enable better connections to universal culture's for individuals as observed in Indigenous communities spanning from Canadian and Aboriginal Australian communities (Dyson 2004; Mignone et al. 2009; Fogarty and Kral 2011; Madden et al 2012; Taylor 2012; Featherstone 2013). This proves there is value available for Aboriginal Australian persons and communities when ICT projects are executed by sole use of the culture (Chikonzo 2006; Schräpel 2010; Taylor 2012). In the Yarnangu community in Australia young people feel that without the capacity to learn about culture and generate art and music at the media centre the community would be a "Sad one, there'd be nowhere to get learnt. They would just sit around, play cards, sit and watch videos and DVDs at home, might be more break-in and sniffing, ganga. Everyone wants the media" in an interview with the author in the Irruntju community (Featherstone 2013).

Issues with Aboriginal Australian use of ICT's embodying Western values have arisen however, with an increase in Internet usage by younger generations in the communities in social media such as Facebook has generated concern by "Ngaanyatjarra Media chairperson Winnie Woods, an advocate of new technologies, expressed concerns about the potential for a breakdown of kinship rules (pers. comm., Irrunytju Community 29 July 2010): 'We don't want our kids getting learned about the Facebook and getting involved. They might get with the wrong woman or the wrong man.'…

Ngaanyatjarra Media has worked closely with Yarnangu since 2004 to develop culturally appropriate models for introducing and engaging with ICTs" (Featherstone 2013).

Consequently, the holistic investigation and incorporation of Aboriginal Australian culture - language and social structures and values - in designing and used in the functional operation of the system were imperative to ensure the systems deliverables would be achieved and the preservation of Aboriginal Australian culture would be done respectfully and not misappropriated. For these purposes, practices that capture these facets of Aboriginal Australian culture were incorporated into the fieldwork methodologies and in directing the approach to community consultation and interviews and surveys.

## 2.3 The Digital Divide and the Embracement of ICT

Some Aboriginal communities have readily embraced ICT as it's become available to their community despite the long existing digital divide, however with the rollout of Next G – a mobile phone network – and the NBN – an internet provision network aiming to cover 93% of Australians needing coverage – young people are observed texting and engaging in chat rooms (Taylor 2012) making digital technology a part of people's everyday lives in the Aboriginal Australian communities with access to these ICT provisions (Gibson 2009; Mohammad and Yi-Chan 2010; Taylor 2012a; 2012b; Featherstone 2013) with many young people having developed computer literacy and creating media such as films, music, presentations, sharing photos and using internet banking and recreational downloading of music and playing online video games (Featherstone 2013).

The increasing uptake in ICT has occurred parallel to the increase in availability and provision of the ICT services for Aboriginal people to utilise these technologies. Combined with the appropriation of use to self-determined purposes such as creation of media and cultural pieces suggests the factors impeding to the adoption of ICT from Aboriginal Australian communities has not been a rejection of Western values embedded in technology – as seen with the increasing uptake of Facebook amongst indigenous youth in Aboriginal Australian communities – and instead, greatest impediment being the lack of accessibility. As seen in Canada and Australia, as the availability of services and the technology have increased and become more accessible there have been a widespread uptake of those technologies that reflect a relationship suggesting as the supply is available, the demand is created (Dyson 2004; Daly 2005; Perley and O'Donnell 2006). In this Aboriginal Australian communities, the digital divide impacts on Aboriginal Australians have depleted over the past decade (Gibson 2009), in this spirit, Aboriginal residents have undertaken a rapid catch up for ICT ownership and use as its availability has been steadily increased (Taylor 2012) in an extent that would allow academics and anthropologists to state that as stated by Inge Kral (2010) 'Aboriginal youth are now firmly part of the new "digital culture".

As the embracement of ICT has increased, so too has the interest from elders in Aboriginal Australian communities to use the ICT to preserve, teach and share cultural knowledge and stories. The most renown approach to this preservation and creation of digital archiving of culture is the Ara Irititja Archives managed by the Anangu people (South Australians Museum) which is used to digitally archive art, crafts, photographs, traditional objects, manuscripts, films, sound recordings and journals and has granted full access to the cultural materials to people which had been previously locked away in state museums and libraries. The information system is designed to be navigated as a virtual throwing of rocks which is the similar custom to how Anangu announced their arrival to a new location (Dyson 2004).

Multimedia preserving platforms have been sufficiently preserving respective Aboriginal Australian cultures for many years, it is no new concept. However, the fault with many of these systems is the inaccessibility via the internet which would otherwise add multitudes of value in the successfulness of the systems purpose – cultural preservation.

Reducing the distances between homes and community centres, which are run through usually small centres and there being access to between one and five computers at each centre makes the process of ascertaining cultural knowledge difficult with demand out measuring supply - the innovative step to utilise the internet as a tool to preserve culture, language and knowledge becomes apparent.

With the increase in personal ownership of ICT's and telecommunications companies providing internet infrastructure to these communities; the valuability of establishing a web based approach is undeniably functional. Additionally, as more people from Aboriginal Australian communities travel for work, education and future opportunities and live in far distanced cities and towns the internet would allow those individuals to remain in touch with their communities culture and not lose it in their travels to other places and to also ensure that individuals who live far from home temporarily or permanently may continue to grow their cultural understanding and have an ongoing sense of their Aboriginality as part of their identity.

| Database Name | Managed | Location/s | Data Types |
|---|---|---|---|
| Ara Irititja | Centrally | 14 sites across the APY lands. A number of other commercial clients across Australia and internationally. | Text, Audio, Video, Image. |
| Our Story | Locally | 14 sites within the Northern Territory. | Text, Audio, Video, Image. |
| Mukurtu | Locally | Tennant Creek, NT. | Text, Audio, Video, Image. |
| CMS | Locally | Uluru, Wet Tropics QLD, Jawoyn Association, NT. | Geospatial, Text, Audio, Video, Image, Rockart data. |
| Traditional Knowledge Recording Project | Locally | Cape York, QLD. | Text, Audio, Video, Image. |
| Martu Kanyirninpa Jukurrpa (Ara Irititja Software) | Centrally | Parnngurr, WA | Text, Audio, Video, Image. |

*Table 1: Currently existing Aboriginal Australian preservation information systems (Gibson 2009)*

## 2.4 ICT Uptake and Engagement with the International Community

Aboriginal Australian communities in remote communities view images of faraway cities and towns and have fostered interest in travel by using the Internet – for worldly experience, for education, for work – and piqued increased migration rates of members of these remote communities (Mignone 2009; Taylor 2012a; 2012b).

This behavioural transition is consequential of increases in social capital in these remote communities. The use of ICT in these remote communities has increased Aboriginal Australians' exposure to the global community (Mignone 2009). To maintain social structures and kinship rules in a population interested in travel and proving higher levels of migration a system which is accessible worldwide to enable Aboriginal Australian's to engage with and not forget cultural practices is imperative.

Web based ICT is an invaluable tool to ensure Aboriginal persons stay in contact with family members (their kinship) as community members embark upon national and international travels and outside proximity with their community's culture. ICT allows communications and engagement with cultures across vast distances and would therefore be extremely useful for the growing number of Aborigines who resolve to travel outside of their communities to maintain cultural sovereignty and engagement. Consequently, the outcomes of culturally aware, community consulted design processes will enable a variety of benefits for Aborigines as far as preservation, learning, and engagement with culture over vast distances is concerned. Consultation processes used for the design and development of these IS's accompanied with support from executive level social members with the appropriate resources to enable the development of such a system will see the successful eventuation of a culturally appropriate, useful system for Aboriginal Australian communities and local land councils to use to remain in touch with culture.

## 3 Research Questions

This paper addresses two (2) main research questions connecting ICT as a preservation tool and Aboriginal Australian culture and language. The use of Activity Theory as a framework to derive methodology – community consultation (see methodology – Community Consultation) – enables the greatest in depth understanding of the Aboriginal Australian culture and how it may interact with ICT, additionally, Activity Theory provides the greatest contextualisation of Aboriginal Australian culture, which will enable the greatest development of appropriate methodologies to best answer the following research questions:

1) How does ICT enable the preservation and immortalisation Aboriginal Australian culture and language?

2) What Information System Development approaches are most appropriate for preserving Aboriginal Australian culture and language, considering literacy proficiency and IT skill levels in Australian Indigenous communities?

    2a) How will Information Systems enable superior preservation of Aboriginal Australian language and culture in comparison to alternative preservation provisions?

    2b) Why prior Aboriginal Australian language and cultural ICT preservation projects/systems failed to preserve or could better preserve Aboriginal Australian culture and language in the past?

Despite Aboriginal Australian archival systems having already existed for years, they have either failed or not realised their true potential. One of such a system is the Koorie Archival System, which, even with strong community commitment, as many archival systems have, has not met its true potential as a preservation tool for Aboriginal Australian communities (Denison et al. 2012). Thus, this paper proposes NSW ALC support for the development and roll out of the system, and individualistic community consultation with the use of Activity Theory to design a sustaining form of knowledge for all Aboriginal Australian LALC's in NSW. The framework these design and development ideas originate from are embodiments of the Activity Theory Framework.

## 3.1 Activity Theory

The activity theory framework will serve as a research lens and will be purposed for consultation with communities and deriving knowledge from collated communal information to design and develop the information system as seen below:

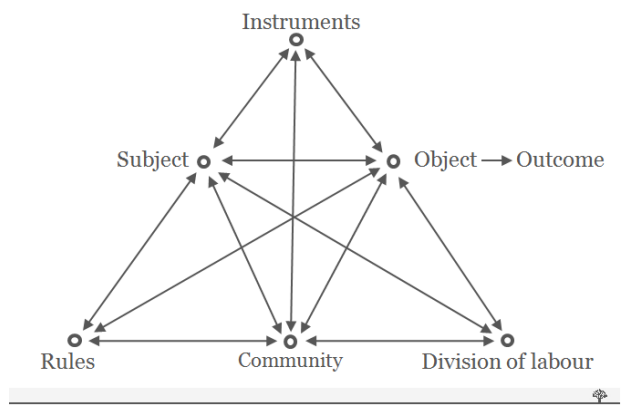

*Figure 1: The Engeström (1987) Activity System.*

These **instruments** used have traditionally been books, paintings, rock drawings, tools and other traditional artifacts, to capture this information the use of ICT - voice recorders, video recorders, and online surveys will contribute to compiling data and information and corresponding mediums these ICT's capture in will be metadata types for the information system.

The **subject** will be the people in the communities; the community members of the LALC's whom will be consulted and surveyed to collect data and structure the metadata existing within the information system within that LALC's database.

The **rules** consist of LALC/ Indigenous laws and traditional rules that impact the sharing and passing of knowledge and interactions between members from communities such as: moiety, kinship, skin colour, totems. These traditional laws and rules will be captured in community consultation and surveys to ensure the correct meta-data format is created to guarantee the capturing of this information is done respectfully to traditional ways.

**Community** refers to the LALC's that will opt in to the development of a system to preserve and teach knowledge.

**Division of labour** is in reference to the segregation of knowledge and teaching between men's business and women's business and that women and men shall not teach one another the segregated gendered business' of the community and other communities. In the process of consultation, additional division of labour metrics will be enacted on an ad hoc basis as mentioned by members of the community and integrated into rules and the subjects.

Finally, the **object** (the outcome) will be to deliver a fully functioning information system with end user driven design and development which conforms to the Indigenous values, rules, and instruments surrounding their communities' knowledge generation and education of knowledge.

This can be understood as one large database which feeds data into LALC specific databases/ servers, through which, the internet can be accessed by members of their respective LALC's. Culturally respective interfaces/ designs accepted by members of the community will ensure the use of the ICT is an enhancer for culture, not a hindrance. User access and system capabilities will abide by the rules of the community and traditionally set divisions of labour between subjects (actors/ users – mens business and womens business). Finally, the use of ICT instruments to access the traditional artefact instruments, which create the object, which is the final, operating, information system.

## 4　Methods for Designing the System

To ensure the information system will archive and can be used to teach cultural knowledge in culturally appropriate manners; manners which will divide Aboriginal Australian knowledge by local council's – regional knowledge - and women's business and men's business to ensure there is not breach in the information that community members receive which will ensure the integrity of the taught culture is upheld and the appropriateness of knowledge is respectful and follows social guidelines that originate from the traditional forms of which knowledge was taught.

To ensure these deliverables would not be harming to the Aboriginal Australian culture, the traditional processes of teaching and sharing knowledge and the community structures and individuals who make up these communities. To ensure these deliverables a down up approach to designing the system must take place and for the development of the system, the resources should be accessed from a top down approach.

In order to correctly navigate the bottom up approach of design and top down approach for development of the system – a thorough understanding of how contemporary Aboriginal Australian social structures operate is imperative. The system should also have cultural influence in the interface for the end users to engage with their culture and to make it representative of traditional teaching and learning of knowledge.

Consequently, a detailed mapping of how local land councils, regional land councils and state land councils interact and interrelate with each other becomes imperative. As seen in fig 2, the state land council acts as what could be analogous in IT projects as executive members who may provide resources and commitment to see the project's success to meet some outcome or deliverable – such as the immortalization and prevention of further loss of Aboriginal culture which is priceless and incredibly important, and the ultimate objective of this project.

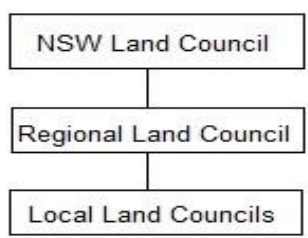

*Figure 2: The hierarchy of the Aboriginal land councils within NSW borders in Australia. Outline of the contemporary Indigenous Australian social structures in place*

The aims and objectives of all land councils at all levels are:

- Land acquisition either by land claim or purchase,
- Establishment of commercial enterprises and community benefit schemes to create a sustainable economic base for Aboriginal communities, and
- Maintenance and enhancement of Aboriginal culture, identity and heritage (including the management of traditional sites and cultural materials within NSW)." (Anonymous 2015b)

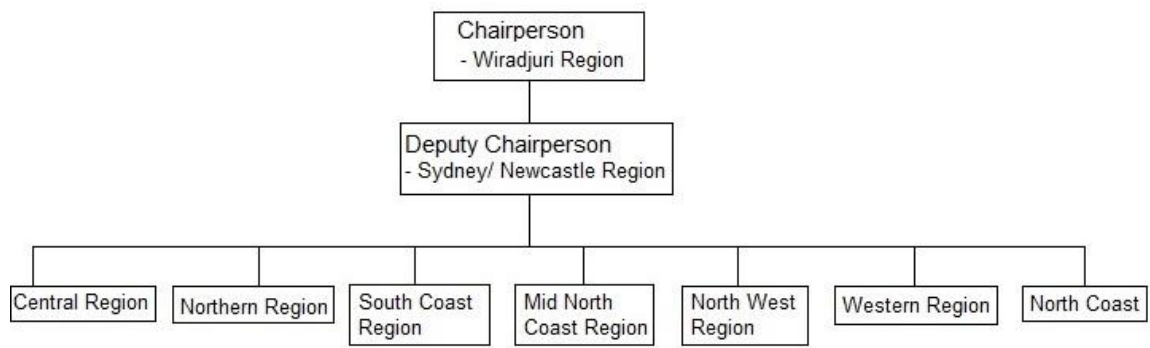

*Figure 3: The structure of the NSW Aboriginal Land Council (NSW ALC) as of May 2015*

The NSW Land Council produces an annual report which provides a very transparent insight into the functions and activities the NSW land council performs on a yearly basis inclusive of events and initiatives performed in the best interest of Aboriginal Australian's. The report includes detailed information about the organization as a whole, inclusive of budget statements. The NSW Aboriginal Land Council (see Figure 4) is comprised of members who are from different Regional Aboriginal Land Councils (RALC's) which represent their Local Aboriginal Land Councils (LALC's).

Each member of the council, inclusive of the chairperson and deputy chairperson represent a region of NSW and advocate priorities and issues in their region which is vocalized by the multiple local land councils in each region. This enables communication from the bottom up through this contemporary social structure which will prove very important for any project that effects the communities in the Local land councils.

The RALC's are composed of multiple members from Aboriginal Local Land Councils that are within the borders of that region. Fig 5. details the current structure as of May 2015 of all of the RLC's in NSW. There are nine regions of which all of the 119 Local Aboriginal Land Councils in NSW's are divided into (http://www.alc.org.au/land-councils/lalc-regions--boundaries.aspx) – the North Western, Northern, South Coast, North Coast, Western, Wiradjuri, Central, Mid North Coast, Sydney/ Newcastle – 9 regions with one councilor from each region – inclusive of chairpersons of the NSW ALC to provide equal representation of different region in state matters.

| North Western | Northern | South Coast | North Coast | Western | Wiradjuri | Central | Mid North Coast | Sydney/Newcastle |
|---|---|---|---|---|---|---|---|---|
| Baradine | Amaroo | Batemans Bay | Baryulgil Square | Balranald | Albury & District | Dubbo | Birpai | Awabakal |
| Brewarrina | Anaiwan | Bega | Birrigan Gargle | Broken Hill | Bathurst | Gilgandra | Bowraville | Bahtabah |
| Collarenebri | Armidale | Bodalla | Bogal | Cobar | Brungle - Tumut | Mudgee | Bunyah | Biraban |
| Coonamble | Ashford | Cobowra | Casino-Boolangle | Dareto | Condobolin | Narromine | Coffs Harbour | Darkinjung |
| Goodooga | Coonabarabran | Eden | Grafton-Ngerrie | Ivanhoe | Cowra | Nyngan | Forster | Deerubbin |
| Lightning Ridge | Dorrigo Plateau | Illawarra | Gugin Gudduba | Menindee | Cummeragunja | Trangie | Karuah | Gandangara |
| Moree | Glen Innes | Jerrinja | Jali | Mutawintji | Deniliquin | Warren Macquarie | Kempsey | La Perouse |
| Mungindi | Guyra Moombahlene | Merriman | Jana Ngalee | Tibooburra | Griffith | Weilwan | Nambucca Heads | Metropolitan |
| Murrawari | Nungaroo | Mogo | Jubullum | Wanaaring | Hay | Wellington | Purfleet/Taree | Mindaribba |
| Narrabri | Red Chief | Ngambri | Muli Muli | Wilcannia | Leeton & District | | Stuart Island | Tharawal |
| Nulla Nulla | Tamworth | Nowra | Ngulingah | Winbar | Moama | | Thungutti | Worimi |
| Pilliga | Walhallow | Ulladulla | Tweed/Byron | | Murrin Bridge | | Unkya | |
| Toomelah | Wanaruah | Wagonga | Yaegl | | Narranderra | | | |
| Walgett | | | | | Onerwal | | | |
| Wee Waa | | | | | Orange | | | |
| Weilmoringle | | | | | Peak Hill | | | |
| | | | | | Pejar | | | |
| | | | | | Wagga Wagga | | | |
| | | | | | Wamba Wamba | | | |
| | | | | | West Wyalong | | | |
| | | | | | Young | | | |

*Figure 4: The 9 Regional Aboriginal Land Councils (RALC's) and the 119 Local Aboriginal Land Councils (LALC's) in each respective region.*

With a thorough understanding of contemporary indigenous social structures, the design and development of the information system can be actualized. The two layered approach to design and development are inversely related and symbiotic of one another.

## Top down Approach: Development and Resources

The top down approach to development of the system involves the NSW ALC providing funding and resources to physically develop the system for each LALC with the confirmation of a LALC opting in and agreeing to have digital archives built in a heavily consulted process to provide usability for the end users. To ensure the system performs as required for each LALC, thorough consultation and input from end users to ensure the system reflects traditional learning within each LALC.

## Bottom up Approach: Design and Consultation

The bottom up approach to design will see end users at the forefront of the system design process to ensure the system reflects the LALC's specific teaching culture and that the LALC's relevant knowledge is archived respectfully and is accessible only for those in that respective LALC. This will enable communities to meet, share stories and knowledge as traditionally done. This will also provide segregation of men's and women's business with user access restrictions. Kinship and moiety knowledge will be accessed as determined per LALC's cultural requirements.

Community consultation would see the system written in appropriate language, creation of an appropriate interface and username and passwords provided to users to ensure the women's and men's business is segregated and women may only access women's business and vice versa. The username and password system would also allow the overarching system to identify which LALC each community member resided within and provide only access to that community's knowledge to ensure community knowledge remained within the community and the sharing of knowledge and stories was performed in traditional ways. As consultation occurs, a deeper understanding of which knowledge's require segregation – perhaps some LALC's require more constraints as to whom may learn certain knowledge that extend beyond the parameters of gender segregation, as these findings occur they will be considered and adopted into the design of the information system for each LALC.

Figures 6 and 7 visually explains the retention of information in one large, overarching system which all LALC's are categorized within and the information related to the LALC resides only within the category of the respective LALC. The data/ information/ knowledge is added to the system by the means of an Aboriginal person signing in – their username will be comprised of their name and their LALC – for example: Sarah Grace from Narrabri would like to create an account so her username would include her name an abbreviation of Narrabri and the two last digits of her D.O.B.. The generated username would be: SarahGNarrabri95. Whilst creating an account to log in to the system, the users will also be asked for their gender, email address (if relevant), secret questions, etc. for account recovery and assurance as to who in which community is logging in to make sure no knowledge is being accessed by individuals external to the community – keeping knowledge secure.

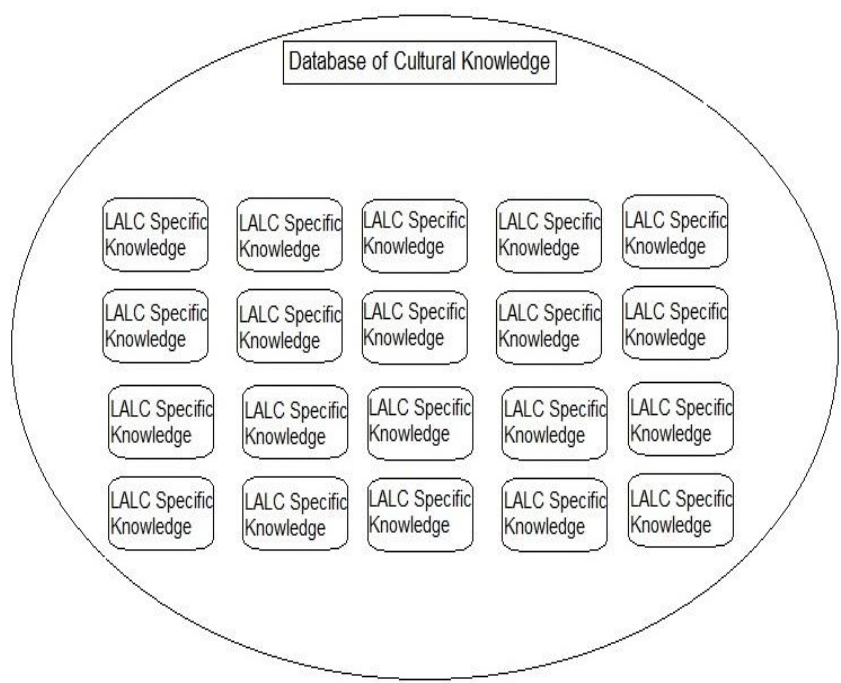

*Figure 5: The proposed system: One large system integrated by NSW ALC, filtered through RALC's, and built/ designed by LALC's to ensure it meets the actor's needs.*

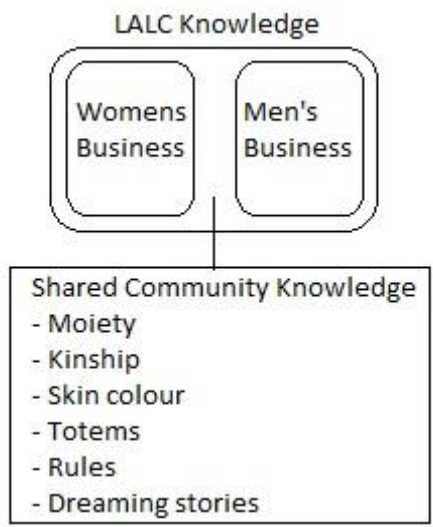

*Figure 6: The separation of men's business and women's business within each LALC will be respected and be integrated into user access controls.*

## Community Consultation

In addition to maintaining integrity for the Aboriginal Australian LALC's and their respective unique knowledge and teaching of that knowledge it is imperative to consult with each LALC to develop the databases and controls as even in contemporary society archived Aboriginal Australian information sources are dispersed and fragmented across governmental departments, private archives and are subject to many forms of custodial agreements (Denison et al., 2012).

The methodologies used to design and develop the information system are derived from the information system development life cycle (SDLC) framework. Respectively, bottom-up, kinship determined access rights, and commitment from the executives of the Aboriginal Land Council social structure (NSW ALC) for the development of the system and for access to resources will determine the methodologies. Through understanding the various needs of each LALC and rules applying to knowledge, community consultation to design the systems bottom up are imperative.

Community consultation will consist of interviews, surveys, meetings, JAD sessions, and recordings of opinions, epistemologies, and feedback will constitute the design and development of the information system through community consultation to ensure the deliverables are integral preservations of the Aboriginal Australian cultures.

## 5   Next Step and Limitations

The next step will be to perform surveys and consultation in Aboriginal Australian communities with the use of ICT's to collect data – video/voice recorders to capture data. The step after this would be the contextualisation of the data in order to generate information about how different knowledge's must be stored. Once understanding how all knowledge needs to be treated with in the system (storage and access rights) the development of a fully functioning, Aboriginal Australian information system to preserve Aboriginal Australian culture and language can exist. After these design and development requirements are determined, the information systems will be built to minimise the loss of cultural knowledge in any immediate capacity.

The limitations will be mass produced generic information systems to display cultural knowledge for all LALC's. Each interface will be catered to each LALC's requirements to be assured that the cultural knowledge is preserved and taught culturally appropriately.

## Acknowledgements

I would like to pay my acknowledgement to Stephen Smith who has guided me in my development to research and write to my capacity as read in this manuscript. I would also like to thank Stephen for his encouragement to pursue all opportunities available.

## Copyright